\documentclass[12pt]{article}
\usepackage{graphicx}

\begin{document}

\title{Tailor Based Allocations for Multiple Authorship: a fractional $gh$-index}
\author{Serge Galam\thanks{serge.galam@polytechnique.edu},\\
Centre de Recherche en \'Epist\'emologie Appliqu\'ee,\\
\'Ecole Polytechnique and CNRS, \\CREA, Boulevard Victor, 32,
75015 Paris, France}

\date{}
\maketitle

\begin{abstract}

A quantitative modification to keep the number of published papers invariant under multiple authorship is suggested. In those cases, fractional allocations are attributed to each co-author with a summation equal to one. These allocations are tailored on the basis of each author contribution. It is denoted ``Tailor Based Allocations (TBA)" for multiple authorship. Several protocols to TBA are suggested. The choice of a specific TBA may vary from one discipline to another. In addition, TBA is applied to the number of citations of a multiple author paper to have also this number conserved. Each author gets only a specific fraction of the total number of citations according to its fractional paper allocation. The equivalent of the $h$-index obtained by using TBA is denoted the $gh$-index. It yields values which differ drastically from those given by the $h$-index. The $gh$-index departs also from $\bar{h}$ recently proposed by Hirsh to account for multiple authorship. Contrary to the $h$-index, the $gh$-index is a function of the total number of citations of each paper. A highly cited paper allows a better allocation for all co-authors while a less cited paper contributes essentially to one or two of the co-authors. The scheme produces a substantial redistribution of the ranking of scientists in terms of quantitative records. A few illustrations are provided. 

\end{abstract}
 

\section*{Introduction}

Individual bibliometry is today a major instrument to allocating research funds, promoting academics and recruiting researchers. The existence of the $h$-index  \cite {h1} has boosted the use of quantitative measures of scientist productions. In particular its incorporation within the Web of Science via a simple button have just turned upside-down the world of evaluation. The use of the $h$-index is now widespread and unavoidable despite all the associated shortcomings and biases.
 
To restrict an individual evaluation to only quantitative figures combining the number of articles, the total number of citations and the $h$-index allow at a glance to rank two competing scientists within a given field. Nevertheless qualitative evaluation is still of a considerable importance to approach a scientist career. 

It is also worth to emphasize that in addition to the institutional use of these indexes and numbers, watching at one's own $h$-index as well as those of friends or competitors has became a ludic and convivial game to place a researcher in the social perspective of its community out of it absorbing lonely state of doing research. 

As expected for any index, the $h$-index was shown to exhibit a series of shortcomings and weaknesses prompting a series of modifications like a weighting by citation impact  \cite{hw} and the bursting of novel proposals with the $g$-index  \cite{g1}. For a review focusing on the the $h$-index variants, computation and standardization for different scientific fields  see \cite{hr} and \cite{hc1} for a comparison with standard bibliometric indicators.  Complements to the h-index \cite{hh, hu} as well as generalizations \cite{hg1} and variants \cite{hg2} of both the $h$ and $g$-indexes have proposed. A comparison of nine different variants of the $h$-index using data from biomedicine has been conducted in \cite{hc2}.

On this basis it is of importance to emphasize that there exists no ultimate index to be self-sufficient. Only combining different indexes could approach a fair and appropriate evaluation of the scientific output of a researcher. 

However, the question of multiple co-authorship has not been given too much of interest although several suggestions have been made recently. For instance it was suggested to rescale a scientist $h$-index dividing it by the mean number of authors of all its papers which belong its $h$-list \cite{batista}. Combining citations and ranking of papers in a fractional way was also proposed \cite{egghe} as well as a scheme to allocate partial credit to each co-author of a paper \cite{hg2}. It was also proposed to count papers fractionally according to the inverse of the number of co-authors \cite{schreiber}.  Last but not least  the initiator of the $h$-index has also proposed to consider a modified $\bar{h}$-index \cite {h2} to account for multiple authors.

It is rather striking to notice that while science is based on the discovering and the use of conservation laws, scientists have been applying the myth of ``Jesus multiplying breads" for decades by multiplying for themselves published articles. Indeed, when an article is published with $k$ authors, each co-author adds one paper to its own list of publications. It means that for one paper published with $k$ authors, $k$ authors add one to their respective number of publications. Accordingly, a single $k$-author papers contributes to $k$ papers while aggregating the total number of papers published by all scientists from their respective publication lists. The same process applies for the citation dynamics with one single citation for one $k$-author paper contributing to an overall of $k$ citations since each one of the $k$ authors includes the citation in its personal $h$-index evaluation.

A few proposal were made previously to conserve the number of articles but prior to the introduction of the $h$-index \cite{zuck, cole, price, cro, vin, van, opp, egghe2} and stayed without much application besides a recent suggestion to define an adapted pure $h$-index  \cite{chai}. Fractional counting was also suggested recently to evaluate universities \cite{loet1, loet2}.

In this paper I propose a new scheme to obey the conservation law of published articles at all levels of associated indexes. One paper counts for a single unit independently of the number of co-authors. This unit must then be divided among the authors. Any fraction used by one of the author is definitively withdraw from the unit. Accordingly, for a multiple author paper fractional allocations are attributed to each co-author with a summation equal to one. These allocations are tailored on the basis of each author contribution. It is denoted ``Tailor Based Allocations (TBA)" for multiple authorship.

The total number of citations given to one paper must be also conserved within the sum of all  credits given to each one of its authors. Any part taken by one author is subtracted from the total. To be consistent the same TBA must be used with respect to all figures attached to a given paper. Each author gets only a specific fraction of the total number of citations according to its fractional paper allocation. Using TBA for citations yields a new equivalent of the $h$-index, I denote the $gh$-index. Contrary to the $h$-index, the $gh$-index is a function of the total number of citations of each paper. A highly cited paper allows automatically a better allocation for all co-authors while a less cited paper contributes essentially either to one or two of the co-authors or little to all authors.

Several protocols to TBA  are suggested and compared. The choice of a specific TBA may vary from one discipline to another. In each case, the $gh$-index yields values which differ drastically from those given by the $h$-index. The $gh$-index departs also from $\bar{h}$ recently proposed by Hirsh to account for multiple authorship. The scheme is found to produce a substantial redistribution of the ranking of scientists in terms of quantitative records. A few illustrations are provided. 

\section*{Designing the perfect author allocations}

According to the principle of conserved number of papers, given a single $k$-authors paper only a fraction  $g(r, k)$ is allocated to each one of the $k$ authors under the constraint 
\begin{equation}
\sum_{r=1}^k g(r, k)= 1  \ ,
\label{w1} 
\end{equation}
where $r=1, 2, ...,k$ denotes the respective position of each author in the sequence of co-authors. The respective values of the set of $\{g(r, k)\}$  are determined following a ``Tailor Based Allocations", the TBA.

All quantitative figures attached to an author at position $r$ of a given $k$ authors paper must then be scaled by $g(r, k)$. Accordingly, the total number of publications of a researcher must be calculated adding the series of the respective fractional  TBA  for all the papers it authored.  Henceforth one paper does not count for one any longer unless it is authored by a single scientist. Given an author with a list of $T$ publications, its total number of articles becomes 
\begin{equation}
T_g  \equiv \sum_{i=1}^T g_i(r, k) \ ,
\label{w2} 
\end{equation}
instead of $T$ with the property $T_g\leq T$.

Similarly, considering the total number of citations of an author, the same rescaling applies. Given a number $n_i$ of citations for paper $i$ in the author list of T publications, the TBA for the paper citations is  
\begin{equation}
g_i(r, k) n_i ,
\label{wgn} 
\end{equation}
instead of $n_i$. Each co-author is granted a different number of citations from the same paper with for a given paper,
\begin{equation}
\sum_{r=1}^k \left[ g_i(r, k) n_i\right]  =n_i \ ,
\label{wn} 
\end{equation}
using Eq. (\ref{w1}). On this basis, the total number of citations of an author becomes 
\begin{equation}
N_g \equiv \sum_{i=1}^T g_i(r, k) n_i  \ ,
\label{w3} 
\end{equation}
instead of $N\equiv \sum_{i=1}^T n_i$ with the property $N_g\leq N$.

Using Eqs. (\ref{wgn}) produces naturally novel values for the corresponding $h$-index denoted $gh$-index. To implement the procedure the next crucial step is to select a criterium to allocate the various values $g(r, k)$ with $r=1, 2,... , k$ for a specific paper. In principle, this set may vary from one paper to another even for the same value of $k$. 

Clearly, the best scheme will eventually become a specific allowance decided by the authors  themselves for each paper prior to have it submitted. For each name, in addition to the affiliation, a quantitative fraction $g(r, k)$ will be stated to denote the fraction of that peculiar paper to be attributed to author $r$. The distribution of numerical values of a series $g(1,k), g(2,k), ...,g(k,k) $ would thus reflect the precise contribution of each one of the authors determining an exact TBA. That will be the most accurate and fair setting. However, an implementation could start only in the near future. In the mean time, we need to adopt one fixed standard in order to make a practical use of our proposal to treat all existing publication data. However, the choice of a protocol must incorporates the tradition of each discipline in co-signing papers. At this stage, each field should adopt its own TBA.

\section*{How to choose a Tailor Based Allocations?}

In physics, and in particular in condensed matter physics, the smallest team, i.e., a group of two persons is composed of one researcher who has performed most of the technical work while the other one has defined the frame and or the problem. Usually the first one is a junior scientist ($J$ ), either undergraduate or graduate student or a postdoctoral researcher whose supervisor is  the second one, a senior researcher ($S$). The associated pair author sequence is then $J - S$. In case we have additional $(k-2)$ authors $A_r$ with $r=2, 4,..., k-1$, the paper becomes a $k$ author paper.  The corresponding name sequence  follows their respective contributions yielding the series $J - A_2 - A_3 - ... - A_{k-1} - S$. However, in terms of decreasing weights of their respective contributions most often we have $A_1- A_k - A_2 - A_3 - ... - A_{k-1}$ for $k$ authors with $J=A_1$ and $S=A_k$, which is different form the name sequence put in the paper. 

While Hirsh advocates a specific scientific policy incentive in designing the $\bar{h}$-index \cite {h2} to favor senior researchers, I focus on trying to incorporate into the ranking of authors the reality of the production of papers. The question of what part of return should be attributed to each contribution is open to a future debate within each discipline to set a standard. The standard could vary from one discipline to another. I consider that in the making of current research the ``technical part" is the one to receive the larger slice of the output. Simultaneously, the ``conceiving part" should be granted with the second larger slice. It follows somehow the spirit of the financial setting of the American National Science Foundation grant attribution. There, the grant pays the full salary of the researcher who does the work against a summer salary for the leading researcher of the grant. That is somehow how it works at least in condensed matter  physics, the field I am familiar with.

I do not intend in promoting one specific policy to favor or discourage conducting collaboration but to build a frame to both capture the current practice and  to exhibit some flexibility to allow adaptation to fit different policies. Various protocols should be first tested in different fields by different researchers, and then it  will become possible to elaborate a standard, which may differ from one field to another. But everyone will thus get its due within the conservation law of published papers applying the TBA. Last but not least, any choice will have the effect to discourage the current inflation of multiple authorship, which automatically increases the ranking of involved scientists. With any TBA, adding an author to a paper will have a ``cost" paid by the others, and in particular to the one who in the current situation is getting credit without doing much work.

\section*{Homogeneous versus heterogeneous TBA}

At this stage to implement our scheme we need to determine an explicit TBA associated to a sequence of authors $A_1- A_2 - A_3 - A_4 - ... - A_{k-1} - A_k$. Previous schemes which did conserve allocation are uniforms with respect to ranking as illustrated with the following three main cases:

\begin{itemize}
\item The simplest equalitarian fractional allocating \cite {price, opp} where each of the $k$ authors receives an allocation $1/k$. However if the output of a paper is equalitarian the input is not making this scheme rather unfair for the author who did the main part of the work.

\item The arithmetic allocating \cite {van} sounds well-balanced with $g(r, k)=\frac{2(k+1-r)}{k(k+1)}$. The higher part  is allocated to first author with $g(1,k)=\frac{2}{(k+1)}$, last author receiving the smaller part $g(k,k)=\frac{2}{k(k+1)}$. This scheme favors the first author at the expense of the last one. In other words, most credit is given to the junior scientist as shown in Figure (\ref{ari}). We could conceive the opposite arithmetic allocation with $g(r, k)=\frac{2(r)}{k(k+1)}$ to heavily benefit to senior researchers as wished by Hirsch. My stand is to favor junior researchers not because they are young but because they do most of the work.

\begin{figure}
\includegraphics[width=1\textwidth]{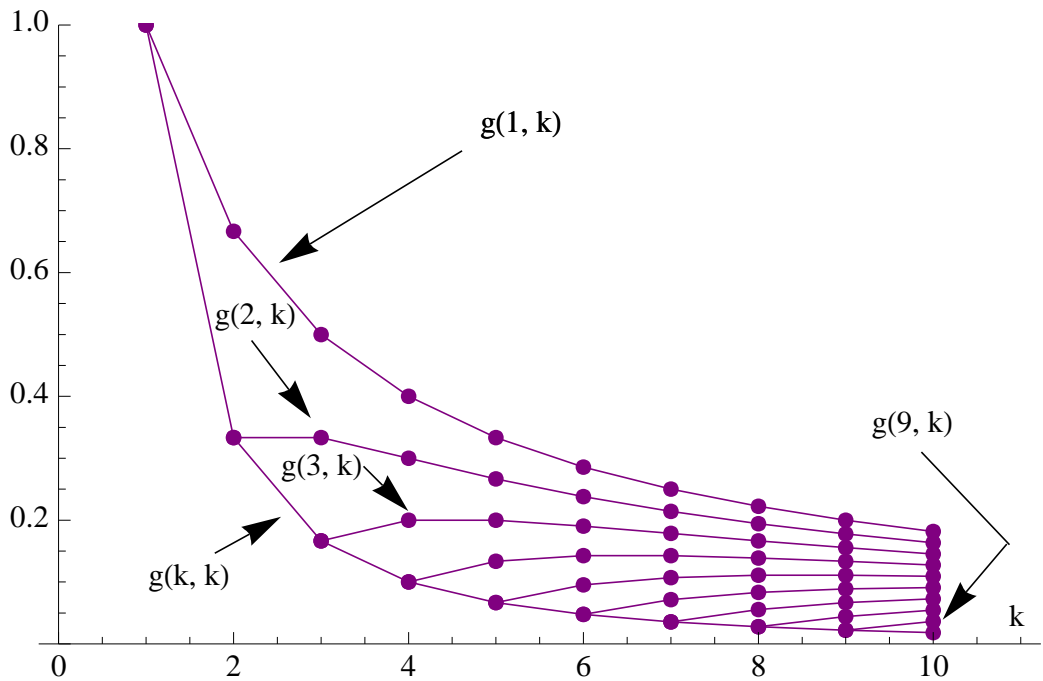}
\caption{The various $g(r, k)$ in case of a decreasing arithmetic allocation to favor junior scientists at the expense of senior ones. The function $g_{k, k}$ represents the slot allocated to the last author of the list. This TBA yields two third one third at $k=2$.}
\label{ari}
\end{figure} 

\item Similarly one can consider a geometric allocating \cite {egghe2} with $g(r, k)=\frac{2^{1-r}}{2(1-2^-k)}$. The higher part  is still allocated to first author with $g(1,k)=\frac{1}{2(1-2^-k)}$, last author receiving the smaller part $g(k,k)=\frac{1}{2^k-1}$.  Figure (\ref{geo}) illustrates the variation of $g(r, k)$ a s function of $k$.

\begin{figure}
\includegraphics[width=1\textwidth]{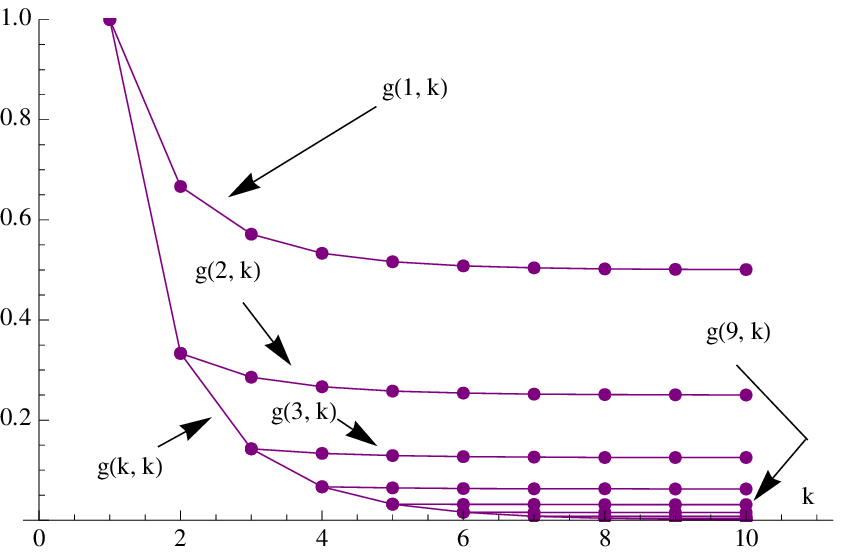}
\caption{The various $g(r, k)$ in case of a decreasing geometric allocation to favor junior scientists at the expense of senior ones. The function $g(k, k)$ represents the slot allocated to the last author of the list. This TBA yields two third one third at $k=2$.}
\label{geo}
\end{figure} 

\item An earlier proposal was a rather awkward combination of equalitarian fractioning  where each one of the authors receives the same slot  $\frac{1}{2(k-1)}$ besides the last author, usually the senior researcher, who gets $\frac{1}{2}$ \cite {zuck}.
\end{itemize}

While above TBA are homogeneous with respect to ranking we now propose an heterogeneous TBA to favor, although differently, both the junior and the senior scientists.
I suggest to allocate extra bonuses, $\delta$ to first and $\mu$ to last authors in addition to the use of a modified non-linear arithmetic allocation to the other authors. Considering $k$ co-authors the formulas are obtained starting from a decreasing arithmetic series $k, k-1, k-2, ..., 2, 1$. The first value $k$ is attributed to the first author to which a bonus $\delta$ is added. The second value $(k-1)$ is given to the last author with a bonus $\mu$. The remaining terms of the series $k-2, ..., 2, 1$ are allocated respectively to authors number $2, 3, ..., k-1$. The sum of all terms yields $S_k= \frac{k(1+k)}{2}+\delta +\mu$, which allows to explicit the various fractional allocations as
\begin{eqnarray}
g(1, k) &= & \frac{k +\delta }{S_k}  \ ,
\label{g1} 
\\
g(k, k)& =&  \frac{k -1+\mu }{S_k}   \ ,
\label{gk} 
\\
g(r, k)&= & \frac{k-r }{S_k}   \ ,
\label{gr} 
\end{eqnarray}
where $g(1, k)$ and $g(k, k)$ are defined only for $k\geq 2$ and $g(r, k)$ only for $k\geq 3$ with $r=2, 3, ..., k-1$.

Next step is to choose the values of the extra bonuses $\delta$ and $\mu$. One hint is to have them determined from setting the case of two authors. At $k=2$ Eqs. (\ref{g1}) and (\ref{gk}) yield $g(1, 2) =\frac{2 +\delta }{S_2}$  and  $g(2, 2)=\frac{1+\mu }{S_2} $ with  $S_2=3+\delta+\mu$. For the case of two author three choices of allocations appear quite naturally with either two to one third, three to one quarter or one to one half. First case is achieved under the constraint $\delta =2 \mu$, second one with $\delta =1+3 \mu$ and the last one with $\delta =-1+ \mu$. Imposing the $k=2$ TBA leaves one degree of freedom for the choice of the overall part attributed to the bonuses when $k>2$. Let us now compare these various choices.

\begin{itemize}

\item{The case ``two to one third"} 

Taking $\delta =2 \mu$ yields $g(1,2)=2/3$ and  $g(2,2)=1/3$. Table (\ref{third}) exhibits all $g(r, k)$ when  $\delta =2$ and $\mu=1$ for $1\leq k \leq 10$ with $1\leq r\leq k$.
For each value of $k$ from 1 to 10 a line gives the various weights $g(r, k)$ calculated from Eqs. (\ref{g1},  \ref{gk}, \ref{gr}) for $1\leq r \leq k$. 

\begin{table}
\centering
\caption{ The various $g(r, k)$ when  $\delta =2$ and $\mu=1$  for $1\leq k \leq 10$ with $1\leq r\leq k$.  At $k=2$ we have two third for the first author and one third for the second one.}
\label{third}   
\begin{tabular}{|c||c|c|c|c|c|c|c|c|c|c|}
\noalign{\smallskip}
k / r &  1 &2 &  3& 4 & 5& 6 & 7 & 8 & 9 &  10 \\ \hline \hline
1  & 1 &  xxx & xxx & xxx & xxx & xxx & xxx & xxx & xxx & xxx     \\ \hline
2  & 0.67 & 0.33 & xxx & xxx & xxx & xxx & xxx & xxx & xxx & xxx   \\ \hline
3  & 0.56 & 0.11 & 0.33 & xxx & xxx & xxx & xxx & xxx & xxx & xxx \\ \hline
4  & 0.46 & 0.15 & 0.08 & 0.31 & xxx & xxx & xxx & xxx & xxx & xxx \\ \hline
5  & 0.39 & 0.17 & 0.11 & 0.05 & 0.28 & xxx & xxx & xxx & xxx & xxx  \\ \hline
6  & 0.33 & 0.17 & 0.12 & 0.08 & 0.04 & 0.25  & xxx & xxx & xxx & xxx \\ \hline
7  & 0.29 & 0.16 & 0.13 & 0.10& 0.06 & 0.03 & 0.23 & xxx & xxx & xxx  \\ \hline
8  & 0.26 & 0.15 & 0.13 & 0.10 & 0.08 & 0.05 & 0.02 & 0.20 & xxx & xxx \\ \hline
9  & 0.23 & 0.15 & 0.12 & 0.10 & 0.08 & 0.06 & 0.04& 0.02 & 0.19 & xxx \\ \hline
10 & 0.21 & 0.14 & 0.12 & 0.10 & 0.09 & 0.07 & 0.05 & 0.03 & 0.02 & 0.17 \\ \hline
\end{tabular}
\end{table}

To visualize the variation of each $g(r, k)$ as a function of $k$ as reported in Table  (\ref{third}), the values are plotted in Figure  (\ref{g23}). Last author of the list received  $g_{k,k}$. It yields two third one third at $k=2$. Every weight $g(r, k)$ starts from $k=r+1$.

\begin{figure}
\includegraphics[width=1\textwidth]{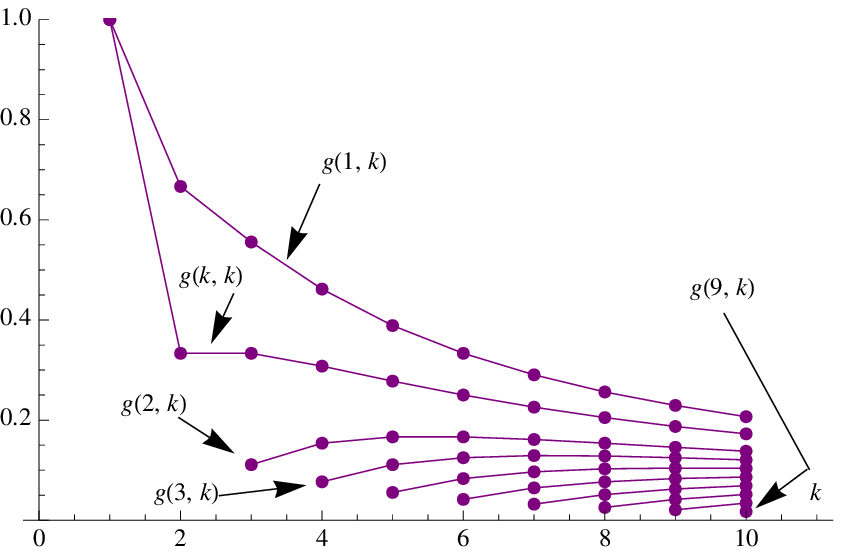}
\caption{The various $g(r, k)$ when  $\delta =2$ and $\mu=1$ for $1\leq k \leq 10$ with $1\leq r\leq k$ from Table  (\ref{third}). The function $g(1,k)$ represents the slot allocated to the first author of the list and $g(k,k)$ the slot allocated to the last author. The bonuses are defined to yield respective allocations of two third and one third at $k=2$. Given a value $r$ the weight $g(r, k)$ is defined starting from $k=r+1$.}
\label{g23}
\end{figure}


\item{The ``case three to one quarter"}

The same as in Subsection \ref{2/3} but with $\delta =1+3 \mu$ to ensure the $k=2$ repartition $g_{1,2} =\frac{2 +\delta }{S_2}=3/4=0.75$  and  $g_{2,2}=\frac{1+\mu }{S_2}=1/4=0.25 $ with  $S_2=3+\delta+\mu$. We select $\delta =1$ and $\mu=0$. The various set of  $g(r, k)$ are listed in Table (\ref{quarter}) and plotted in Figure (\ref{g34}).

\begin{table}
\centering
\caption{The various $g(r, k)$ when  $\delta =1$ and $\mu=0$ for $1\leq k \leq 10$ with $1\leq r\leq k$. At $k=2$ it yields three to one quarter for respectively the first and second authors.}
\label{quarter}   
\begin{tabular}{|c||c|c|c|c|c|c|c|c|c|c|}
\noalign{\smallskip}
k / r&  1 &2 &  3& 4 & 5& 6 & 7 & 8 & 9 &  10  \\ \hline \hline
1  & 1 &  xxx & xxx & xxx & xxx & xxx & xxx & xxx & xxx & xxx     \\ \hline
2  & 0.75 & 0.25 & xxx & xxx & xxx & xxx & xxx & xxx & xxx & xxx \\ \hline
3  & 0.57 & 0.14 & 0.29 & xxx & xxx & xxx & xxx & xxx & xxx & xxx \\ \hline
4  & 0.45 & 0.18 & 0.09 & 0.27 & xxx & xxx & xxx & xxx & xxx & xxx \\ \hline
5  & 0.37 & 0.19 & 0.12 & 0.06 & 0.25 & xxx & xxx & xxx & xxx & xxx  \\ \hline
6  & 0.32 & 0.18 & 0.14 & 0.09 & 0.04 & 0.23 & xxx & xxx & xxx & xxx  \\ \hline
7  & 0.28 & 0.17 & 0.14 & 0.10 & 0.07 & 0.03 & 0.21 & xxx & xxx & xxx  \\ \hline
8  & 0.24 & 0.16 & 0.13 & 0.11 & 0.08 & 0.05 & 0.03 & 0.19  & xxx & xxx \\ \hline
9  & 0.22 & 0.15 & 0.13 & 0.11 & 0.09 & 0.06 & 0.04 & 0.02 & 0.17  & xxx \\ \hline
10 & 0.16 & 0.14 & 0.12 & 0.11 & 0.09 & 0.07 & 0.05 & 0.04 & 0.02 & 0.16  \\ \hline
\end{tabular}
\end{table}

\begin{figure}
\includegraphics[width=1\textwidth]{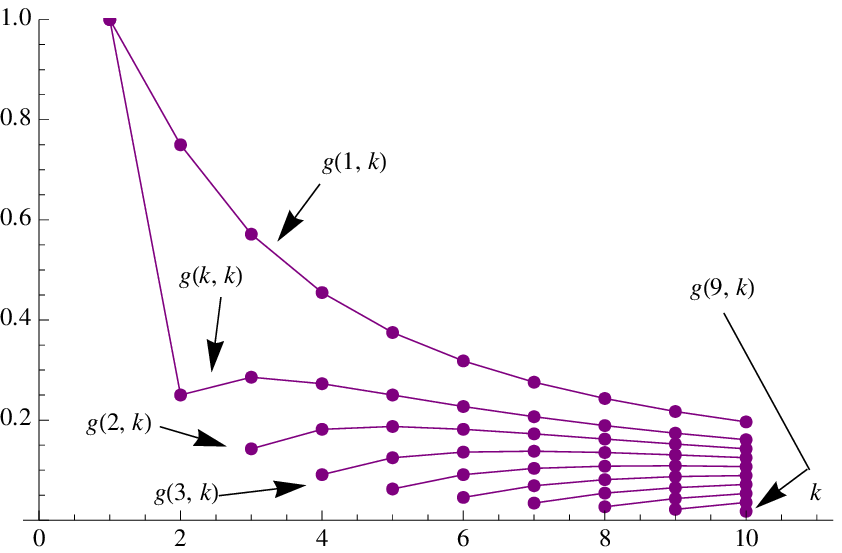}
\caption{The various $g(r, k)$ when $\delta =1$ and $\mu=0$ for $1\leq k \leq 10$ with $1\leq r\leq k$ from Table  (\ref{quarter}). The function $g(1,k)$ represents the slot allocated to the first author of the list and $g(k,k)$ the slot allocated to the last author. The bonuses are defined to yield respective allocations of three to one quarter at $k=2$. Given a value $r$ the weight $g(r, k)$ is defined starting from $k=r+1$.}
\label{g34}
\end{figure} 


\item{The case ``one to one half"}

The same as in Subsection \ref{2/3} but with $\delta =-1+ \mu$ to ensure the $k=2$ repartition $g(1,2) =\frac{2 +\delta }{S_2}=1/2=0.50$  and  $g(2,2)=\frac{1+\mu }{S_2}=1/2=0.50 $ with  $S_2=3+\delta+\mu$. We select $\delta =0$ and $\mu=1$. The various set of $g(r, k)$ are listed in Table (\ref{fifty}) and plotted in Figure (\ref{g12}).
 
\begin{table}
\centering
\caption{The various $g(r, k)$ when  $\delta =0$ and $\mu=1$ for $1\leq k \leq 10$ with $1\leq r\leq k$. At $k=2$ it yields one to one half.}
\label{fifty}   
\begin{tabular}{|c||c|c|c|c|c|c|c|c|c|c|}
\hline\noalign{\smallskip}
k / r &  1 &2 &  3& 4 & 5& 6 & 7 & 8 & 9 &  10 \\ \hline \hline
1  & 1 &  xxx & xxx & xxx & xxx & xxx & xxx & xxx & xxx & xxx \\ \hline 
2  & 0.50 & 0.50 & xxx & xxx & xxx & xxx & xxx & xxx & xxx & xxx  \\ \hline
3  & 0.43 & 0.14 & 0.43 & xxx & xxx & xxx & xxx & xxx & xxx & xxx \\ \hline
4  & 0.36 & 0.18 & 0.09 & 0.36 & xxx & xxx & xxx & xxx & xxx & xxx  \\ \hline
5  & 0.31 & 0.19 & 0.12 & 0.06 & 0.31 & xxx & xxx & xxx & xxx & xxx \\ \hline
6  & 0.27 & 0.18 & 0.14 & 0.09 & 0.04 & 0.27 & xxx & xxx & xxx & xxx \\ \hline
7  & 0.24 & 0.17 & 0.14 & 0.10 & 0.07 & 0.03 & 0.24 & xxx & xxx & xxx  \\ \hline
8  & 0.22 & 0.16 & 0.13 & 0.11 & 0.08 & 0.05 & 0.03 & 0.22 & xxx & xxx  \\ \hline
9  & 0.20 & 0.15 & 0.13 & 0.11 & 0.09 & 0.06 & 0.04 & 0.02 & 0.20 & xxx  \\ \hline
10 & 0.18 & 0.14 & 0.12 & 0.11 & 0.09 & 0.07 & 0.05 & 0.04 & 0.02 & 0.18  \\ \hline
\end{tabular}
\end{table}

\begin{figure}
\includegraphics[width=1\textwidth]{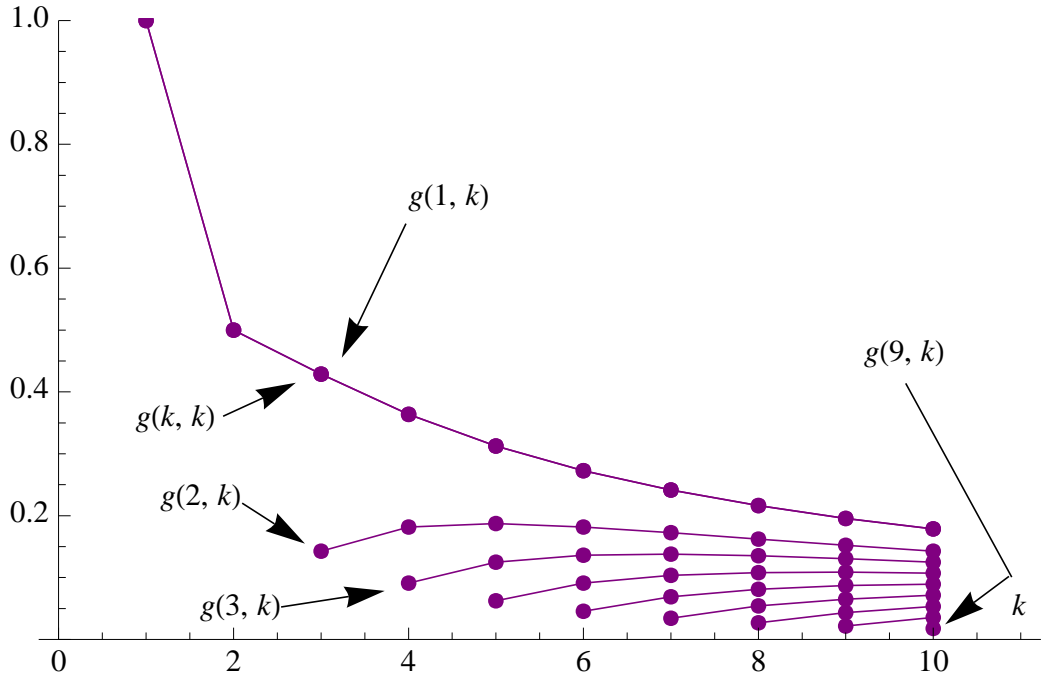}
\caption{The various $g(r, k)$ when $\delta =1$ and $\mu=0$ for $1\leq k \leq 10$ with $1\leq r\leq k$ from Table  (\ref{fifty}). The function $g(1,k)$ represents the slot allocated to the first author of the list and $g(k,k)$ the slot allocated to the last author. The bonuses are defined to yield respective allocations of one to one half at $k=2$. Given a value $r$ the weight $g(r, k)$ is defined starting from $k=r+1$.}
\label{g12}
\end{figure} 


\item{The inhomogeneous arithmetic case}

Putting both bonuses equal to zero $\delta=\mu=0$, Eqs. (\ref{g1}, \ref{gk}, \ref{gr}) recover an arithmetic series where the second term is attributed to the last author.
Associated values are reported in Tables (\ref{ar}) and shown in Figure (\ref{g00}), which is almost identical to Figure  (\ref{ari}) but yet with subtle differences. Here $g(k,k)$ stands instead of $g(2,k)$ and $g(2,k)$ stands instead of $g(3,k)$. In addition $g(9,k)$ at $k=10$ is the last lower point while it is the one before last in Figure  (\ref{ari}). Depending on which situation has been selected either Tables (\ref{ar}) or Tables (\ref{ari}) the strategy of the senior scientist in adding authors will be totally opposite.

\end{itemize}

\begin{table}
\centering
\caption{The various $g(r, k)$ when $\delta=\mu=0$ for $1\leq k \leq 10$ with $1\leq r\leq k$.}
\label{ar}   
\begin{tabular}{|c||c|c|c|c|c|c|c|c|c|c|}
\noalign{\smallskip}
k / r &  1 &2 &  3& 4 & 5& 6 & 7 & 8 & 9 & 10 \\ \hline
1  & 1 &  xxx & xxx & xxx & xxx & xxx & xxx & xxx & xxx & xxx     \\ \hline
2  & 0.67 & 0.33 & xxx & xxx & xxx & xxx & xxx & xxx & xxx & xxx \\ \hline
3  & 0.50 & 0.33 & 0.17 & xxx & xxx & xxx & xxx & xxx & xxx & xxx \\ \hline
4  & 0.40 & 0.30 & 0.20 & 0.10 & xxx & xxx & xxx & xxx & xxx & xxx \\ \hline
5  & 0.33 & 0.27 & 0.20 & 0.13 & 0.07 & xxx & xxx & xxx & xxx & xxx \\ \hline
6  & 0.29 & 0.24 & 0.19 & 0.14 & 0.09 & 0.05 & xxx & xxx & xxx & xxx  \\ \hline
7  & 0.25 & 0.21 & 0.18 & 0.14 & 0.11 & 0.07 & 0.04 & xxx & xxx & xxx \\ \hline
8  & 0.22 & 0.19 & 0.17 & 0.14 & 0.11 & 0.08 & 0.06 & 0.03 & xxx & xxx \\ \hline
9  & 0.20 & 0.18 & 0.16 & 0.13 & 0.11 & 0.09 & 0.07 & 0.04 & 0.02 & xxx \\ \hline
10 & 0.18 & 0.16 & 0.14 & 0.13 & 0.11 & 0.09 & 0.07 & 0.05 & 0.04 & 0.02 \\ \hline
\end{tabular}
\end{table}

\begin{figure}
\includegraphics[width=1\textwidth]{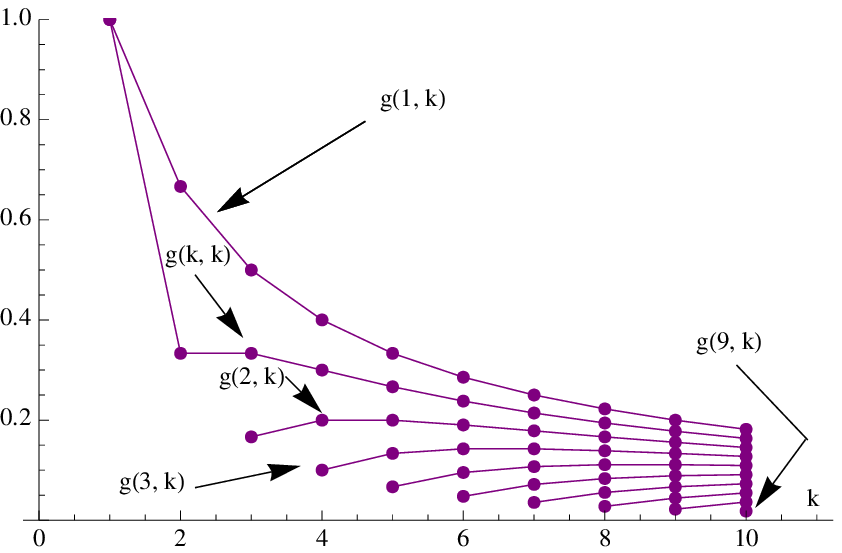}
\caption{The various $g(r, k)$ when $\delta =\mu=0$ for $1\leq k \leq 10$ with $1\leq r\leq k$ from Table  (\ref{ar}). The function $g(1,k)$ represents the slot allocated to the first author of the list and $g(k,k)$ the slot allocated to the last author.}
\label{g00}
\end{figure} 


\section*{Calculating the $gh$-index}

While applying a TBA clearly modifies drastically the number of publications of authors, it also modifies the associated $h$-index by extending the TBA to the paper citations.  For a $k$ author paper, the author at position $r$ in the name sequence now receives $g(r, k) n$ for its own part of citations instead of the total number $n$. Using this fractional number of citations to calculate the $k$-index according to the same definition yields a new lower $gh$-index. The rescaled value does depend on the choice of the bonuses $\delta$ and $\mu$. 

While the choice of these bonuses should be the result of a consensus among researchers, I choose here one arbitrary set to illustrate how the $h$-index is changed. I selected one physicist with a $h$-index equal to 33 to apply my procedure. It names is not disclosed to maintain individual privacy. I also report only its forty first papers of its complete list of articles which is much longer. They are shown in Table (\ref{40-1}) where each paper is ranked with $i=1, 2, ..., 40$ according to its associated total number of citation $n_i$ following a decreasing order. The second column $(r,k)$ indicates respectively the number of authors and the researcher rank of each paper. The last four columns report the rescaled numbers of citations allocated to the author using the precedent four different choices of slot allocations at $k=2$. We have $g(r,k;2/3)$ for the case ``two to one third", $g(r,k;3/4)$ for the case ``three to one quarter", $g(r,k;1/2)$ for the case one to one half and $g(r,k;0)$ for zero bonuses.

\begin{table}
\centering
\caption{An author $h$-index list of publication revisited by four different TBA}
\label{40-1}   
\begin{tabular}{|c|c|c|c|c|c|c|c|c|c|c|c}
\hline\noalign{\smallskip}
$i$ & $ (r,k)$&$n_i$&$g(r,k;2/3)n_i$& $g(r,k;3/4)n_i$& $g(r,k;1/2)n_i$& $g(r,k;0)n_i$\\ 
\noalign{\smallskip}
\hline \hline
\noalign{\smallskip}
1&(2, 2) &187&61.71&46.75&93.5&61.71  \\ \hline
2&(1, 1) &181&181&181&181&181 \\ \hline
3&(3, 3)& 179&59.07&51.91&76.97&30.43 \\ \hline
4&(1, 1)&145&145&145&145&145 \\ \hline
5&(1, 1)&145&145&145&145&145 \\ \hline
6&(2, 3) &132&14.52&18.48&18.48&43.56 \ \\ \hline
7&(1, 1)&132&132&132&132&132 \\ \hline
8&(2, 3) &120&13.20&16.80&16.80&39.60 \\ \hline
9&(1, 2) &104&69.68&78.00&52.00&69.68\\ \hline
10&(3, 1)&98 &54.98&55.86&42.14&49.00   \\ \hline
11&(1, 2) &94&62.98&70.50&47.00&62.98\\ \hline
12&(3, 3) &90&29.70&26.10&38.70&15.30\\ \hline
13&(3, 3) &81& 26.73&23.49&34.83&13.77\\ \hline
14&(1, 1) &75&75&75&75&75\\ \hline
15&(2, 2) &72&23.76&18&36&23.76  \\ \hline
16&(3, 3) &71&23.43&20.59&30.53&12.07\\ \hline
17&(3, 3) &68&22.44&19.72&29.29&11.56\\ \hline
18&(3, 3) &66&21.78&19.14&28.38&11.22 \\\hline
19&(2, 2) &63&20.79&15.75&31.50&20.79\\ \hline
20&(3, 3) &55&18.15&15.95&23.63&9.35\\ \hline
21&(1, 1) &51&51&51&51&51\\ \hline
22&(2, 2) &50&16.5&12.5&25&16.5\\ \hline
23&(2, 2) &48&15.84&12&24&15.84\\ \hline
24&(1, 1) &45&45&45&45&45\\ \hline
25&(1, 1) &43&45&45&45&45\\ \hline
26&(1, 2) &42&28.14&31.50&21.00&28.14 \\ \hline
27&(1, 1) &39&39&39&39&39\\ \hline
28&(2, 3) &38&4.18&5.32&5.32&12.54\\ \hline
29&(2, 2) &38&12.54&9.5&19&12.54 \\ \hline
30&(2, 2) &35&11.55&8.75&17.5&11.55\\ \hline
31&(2, 2) &35&11.55&8.75&17.5&11.55 \\ \hline
32&(2, 2) &34&11.22&8.50&17&11.22    \\ \hline
33&(4, 6) &33&2.64&2.97&2.97&4.62\\ \hline  \hline
34&(1, 2) &31&20.77&23.25&15.50&20.77  \\ \hline
35&(3, 3) &30&9.90&8.70&12.90&5.10\\ \hline
36&(2, 2) &30&9.9&7.5&15&9.9 \\ \hline
37&(3, 3) &30&9.90&8.70&12.90&5.10\\ \hline
38&(2, 2) &30&9.9&7.5&15&9.9 \\ \hline
39&(2, 2) &29&9.57&7.25&14.5&9.57 \\ \hline
40&(2, 2) &29&9.57&7.25&14.5&9.57 \\ \hline

\end{tabular}
\end{table}

Indeed the selected author has published 9 times alone, 4 times as first author for two authors, 13 times as second author for two authors, 1 time as first author for three authors, 3 times as second author for three authors, 9 times as last author for three authors, and 1 time as fourth author for six authors. The associated values of $g(r, k)$ are reported in Table  (\ref{40-2}). They are used to calculate the rescaled citations of last four columns of Table (\ref{40-1}).

\begin{table}
\centering
\caption{ The TBA coefficients used for the author list form Table (\ref{40-1}) }
\label{40-2}   
\begin{tabular}{|c|c|c|c|c|c|c|c|c|c|c|c}
\hline\noalign{\smallskip}
$ (r,k)$&$g(r,k;2/3)$& $g(r,k;3/4)$& $g(r,k;1/2)$& $g(r,k;0)$\\ 
\noalign{\smallskip}
\hline \hline
\noalign{\smallskip}
(1, 1) &1&1&1&1 \\ \hline
(1, 2) &0.67&0.75&0.50&0.67  \\ \hline
(2, 2) &0.33&0.25&0.50&0.33  \\ \hline
(1, 3) &0.56&0.57&0.43&0.50  \\ \hline
(2, 3) &0.11&0.14&0.14&0.33  \\ \hline
(3, 3) &0.33&0.29&0.43&0.17  \\ \hline
(4, 6) &0.08&0.09&0.09&0.14  \\ \hline
\end{tabular}
\end{table}

Using Table (\ref{40-1}) to evaluate the associated $gh$-index, instead of the $h$-index value of 33 we find that  $gh(2/3)=21$, $gh(3/4)=19$, $g(1/2)=23$, $gh(o)=20$. The total number of articles fir the author is respectively 19.91, 18.94, 22.31, 19.13 instead of the inflated value of 40. 

While all choices reduce by approximately half the number of articles, the $h$-index is reduced by one third. The differences between the various sets of $g(r, k)$ are significant but not substantial. Clearly the modifications will vary from one author to another depending on its distribution of multiple collaboration and on the author position in the sequence of names.

\section*{Conclusion}

I have presented a scheme to obey the conservation of both printed papers and given citations. While the principle of a Tailor Based Allocations (TBA) is a scientific prerequisite, the differences between the various sets of fractional coefficients $g(r, k)$ attributed to authors at position $r$ in a list of $k$ names, are significant but not substantial in the cutting of  the current inflation of counting of papers driven by the individual counting of multiple authorship.

More applications are needed to figure out which TBA is more appropriate for each discipline. However, our procedure is readable applicable to any individual set obtained from the Web of Science. 

From our results, it could appear that the TBA rescaling disadvantages senior authors who usually sit last with several co-authors but indeed it is not quite the case since all indexes are deflated to obey the conservation law of existing articles. Moreover, in contrast to the $h$-index, the $gh$-index takes into account the existence of high citations for a paper since then a large $n_i$ does yield large  $g(r, k)n_i$ for all $k$ authors whatever their respective rank is. On the contrary, a low citation paper does contribute mainly to first and last authors.

It is worth to notice that in some disciplines the question of who has contributed the most or least to a paper is uncoupled from the sequence of co-authors in the byline of the paper as traditionally often applied in mathematics and for  practical reasons in high-energy physics. In some cases, it is also known that the ``senior" co-author did not contribute much to the work but cosigns the paper as a privilege of its status being the professor, director, head of department or project leader. These facts make difficult to adopt one single attribution model, which will be fair for all cases of multiple authorship. In those cases, one possibility could be to apply the $1/k$ allocation for $k$ co-authors although the various contributions are rarely equivalent. The situation will be different in the future since then authors will decide which fractions each one gets by having those fractions written along the author affiliations. The sequence of authors could then be at will independent or dependent of the sequence of the authors. 

I do not intend to promote a peculiar policy for collaboration but to set a frame in which one paper counts as one paper independently of its number of authors. On this basis the choice of author allocations should integrate the reality of what part everyone did in the building of a paper with the setting a specific TBA for each paper paper having in mind that one paper counts for one no matter the number of co-authors. The focus of the paper is to show explicitly that applying the TBA to obey the rule that one paper counts to one modifies drastically the ranking of scientists. In particular extending the TBA to the number of citations turns useless to be an author of a multiple author paper with a bibliometric  return of almost zero. The ranking of scientists is also disrupted substantially even if we select the equalitarian $1/k$ TBA for $k$ co-authors  since single and pair authors will be favored with respect to larger sets of co-authors.

My proposal does not aim at setting a specific standard to TBA but to trigger novel practice framed by the constraint of a total weight of one to one paper. In particular, within TBA adding an author to a paper has a ``cost" paid by the others, and in particular to the ones who in the current situation are getting credit without doing much work.

 This principle of reality should not discourage collaborations but on the contrary favor a fair return to multiple authorship. It creates an incentive to stop the current inflation of publications driven by the individual counting of multiple authorship.


\end{document}